\documentclass[preprint, aps, pre, eqsecnum,amsmath, amssymb]{revtex4}
\usepackage[final]{graphicx}

\begin{document}

\title{Magnetization of rotating ferrofluids: the effect of polydispersity}
\author{A.~Leschhorn, J.~P.~Embs, M.~L\"{u}cke}
\affiliation{Institut f\"{u}r Theoretische Physik, Universit\"{a}t
des Saarlandes, D-66041~Saarbr\"{u}cken, Germany\\}

\date{\today}

\begin{abstract}
The influence of polydispersity on the magnetization is analyzed in a
nonequilibrium situation  where a cylindrical ferrofluid
column is enforced to rotate with constant frequency like a rigid body
in a homogeneous magnetic field that is applied
perpendicular to the cylinder axis. Then, the magnetization
and the internal magnetic field are not longer
parallel to each other and their directions differ from that of the
applied magnetic field. Experimental results on the transverse
magnetization component perpendicular to the applied field are compared
and analyzed as functions of rotation frequency and field strength with
different polydisperse Debye models that take into account the
polydispersity in different ways and to a varying degree.
\end{abstract}

\maketitle

\vskip2pc


\section{Introduction}
The prospect of influencing the rotational dynamics of the
nanoscaled magnetic particles in a ferrofluid by a macroscopic flow
and/or by a magnetic field in order to then observe the resulting
response via the magnetization and/or via changes in the flow has
been stimulating many research activities
\cite{Ro85,BlCeMa97,Od02a,Od02b} ever since McTague measured
\cite{McTague69} the so-called magneto-viscous effect. Of particular
interest are in this context flows that are shear free on the
macroscopic scale as in a fluid that is rotating like a rigid body
with a rotation frequency, say, ${\bf \Omega}=\Omega {\bf e}_z$.

While the colloidal magnetic particles then undergo thermally sustained
rotational and translational Brownian motion on the microscopic
scale they co-rotate in the mean with the deterministic macroscopic
rigid body flow. However, this mean co-rotation can be hindered by magnetic
torques on their moments when a magnetic field, say, ${\bf H}_0 =
H_0 {\bf e}_x$ is applied perpendicular to the rotation axis ${\bf
e}_z$ of the flow. The combination of the externally imposed forcing
of the particle motion by (i) the rigid body flow in which they are
floating and by (ii) the magnetic torques on their magnetic moments
drives the colloidal suspension out of equilibrium. Concerning the
magnetic moments, this forcing causes the mean orientation of the
moments, i.e., of the magnetization ${\bf M}$ of the ferrofluid to
be no longer parallel to the internal magnetic field ${\bf H}$.
Instead, ${\bf M}$ is pushed out of the direction of ${\bf H}$ as
well as of that of ${\bf H}_0$ thereby acquiring a nonzero
transverse component $M_y$. Here it should be noted that in a long
cylinder Maxwell's equations imply the vector relation ${\bf H} =
{\bf H}_0 - {\bf M}/2$ between the three fields when they are
stationary and homogeneous but they need not be collinear.
However, in equilibrium, ${\bf \Omega}=0$, the three fields are
indeed collinear: the equilibrium magnetization ${\bf M}_{eq}({\bf
H})=M_{eq}(H){\bf H}/H$ is parallel to the internal field ${\bf H}$.

Recently, the transverse magnetization $M_y$ of a slender cylindrical
column of ferrofluid that was enforced to rotate like a rigid body
with constant frequency $\Omega {\bf e}_z$ in an applied homogeneous
magnetic field $H_0 {\bf e}_x$ was measured as a function of $\Omega
$ and $H_0$ \cite{EmMaWaKiLeLu06}. These measurements showed among
others that the predictions \cite{LeLu06b} based on models for the
magnetization dynamics \cite{Sh72,FeKr99,Fe00,Sh01a,MuLi01} with a
single relaxation time overestimate the magnitude of $M_y$. One
reason for this discrepancy seems to be that particles with
different sizes and different rotational dynamics of their magnetic
moments contribute differently to the non-equilibrium, flow-induced
component $M_y$ of the magnetization. In particular only the
magnetic moments of the larger particles in which the magnetic
moments are blocked and "frozen" in the particles, i.e., those with
effective Brownian relaxation dynamics may be rotated by the flow
out of the direction of the magnetic field.

Here we consider poly-disperse models with single-particle Brownian
as well as N\'eel relaxation dynamics for the different particle
sizes. So we ignore the influence of any dipolar magnetic
interaction and of any flow induced interaction on the (rotation)
dynamics of the particles. Thereby collective, collision dominated
long-range and long-time hydrodynamic relaxation dynamics of the
ensemble of magnetic moments are discarded since only the individual
relaxation of each magnetic moment is considered --- albeit in the
collectively generated internal magnetic field ${\bf H}$.

\section{Equilibrium magnetization}
In our experiments we used several ferrofluids out of the APG-series
of FerroTec. Their saturation magnetization was specified by the
manufacturer to be  $M_{sat}^{FF}=17507$ A/m ($\pm\,10\%$). This
corresponds to a volume concentration $\phi\approx 3.6\%$ of the
suspended magnetic material. We have measured the equilibrium
magnetization of the ferrofluids with a vibrating sample
magnetometer (LakeShore 7300 VSM) with a commercial PC user package.
In order to get information on the particle size distribution of the
ferrofluid under investigation, we used fits
\cite{EmMuKrMeNaMuWiLuHeKn01} with a lognormal form of the
distribution as well as with a regularization procedure
\cite{EmMaWaKiLeLu06} based on Tichonovs method \cite{WeSt85}.
Generally the equilibrium magnetization $M^{eq}(H)$ as a function of
the internal magnetic field $H$ can be approximated by a
superposition of Langevin-functions
\begin{equation}\label{meqvonh}
    M^{eq}(H)=\sum\limits_{j}w_j {\mathcal{L}}[\alpha_j(H)].
\end{equation}
Here ${\mathcal{L}}(x)=\coth(x)-1/x$ denotes the Langevin-function
that depends on the dimensionless Langevin-parameter
$\alpha_j(H)=\mu_0m_jH/k_BT$ and $w_j$ are the so-called magnetic
weights. $m_j$ refers to the magnetic moment of particles with
magnetic diameter $d_j$, i.e.,
$m_j=\frac{\pi}{6}d_j^3M_{sat}^{bulk}$ with $M_{sat}^{bulk}$ the
bulk-saturation magnetization. From Eq. (\ref{meqvonh}) we can
deduce the initial susceptibility $\chi_0=\frac{\pi\mu_0
M_{sat}^{bulk}}{18k_BT}\sum_{j} w_j d_j^3$ and the saturation
magnetization $M_{sat}^{FF}=\sum_j w_j$ of the ferrofluid under
investigation.

\begin{figure}[htp]
\centerline{\includegraphics[width=12cm,angle=0]{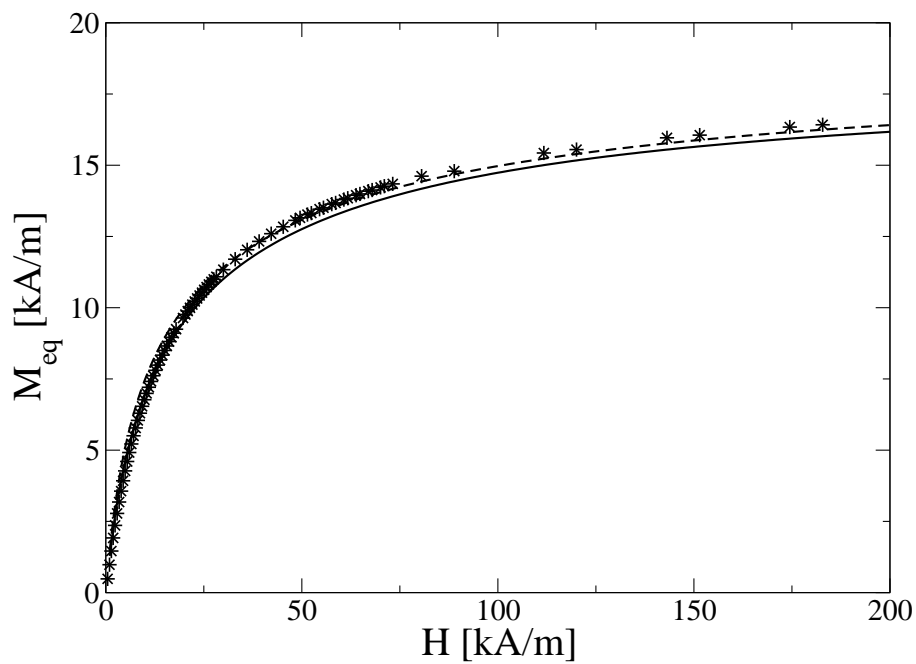}}
\caption{Equilibrium magnetization versus internal magnetic field
for the ferrofluid AGP 933 of FerroTec. Symbols denote experimental
data, solid line fit with a lognormal distribution, and dashed line
fit with a regularization method. See text for further information}
\label{FIG:meq}
\end{figure}

Fig.~\ref{FIG:meq} shows the experimentally determined equilibrium
magnetization $M_{eq}(H)$ of APG 933 versus internal field $H$
together with fits that were obtained with a lognormal distribution
\cite{EmMuKrMeNaMuWiLuHeKn01} and with the regularization method
\cite{EmMaWaKiLeLu06}. The saturation magnetization of the
ferrofluid sample was $M_{sat}^{FF}=19108.6$ A/m. From the
saturation magnetization the volume concentration of the magnetite
particles was found to be $\phi=M_{sat}^{FF}/M_{sat}^{bulk}=4.1$ \%,
in reasonable agreement with the manufacturer's specifications. For
the initial susceptibility we found the value $\chi_0=1.09$.

The magnetic weight distributions $w(d)$ resulting from the two fit
methods are shown in Fig.~\ref{FIG:vert}.

\section{Experimental setup}
\begin{figure}[htp]
\centerline{\includegraphics[width=10cm,angle=0]{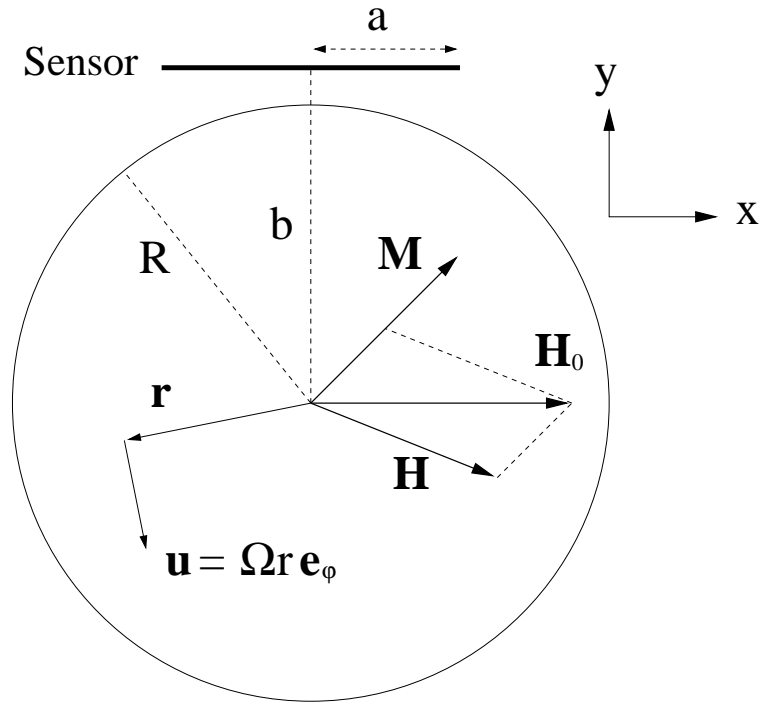}}
\caption{Schematic plot of the system. The cylindrical sample holder
with inner radius $R$ rotates with angular velocity
$\boldsymbol{\Omega}$ in the applied static magnetic field
${\bf{H}}_0$ perpendicular to $\boldsymbol{\Omega}$. The
magnetization $M_y$ is measured with a Hall sensor.  ${\bf M}$ and
${\bf H}$ denote the magnetization and internal magnetic field of
the ferrofluid. Both are constant in space and time.}
\label{FIG:sys}
\end{figure}

The experimental setup for measuring the magnetization of a rotating
cylindrical column of ferrofluid is sketched in Fig. \ref{FIG:sys}.
It is described in more detail in \cite{EmMaWaKiLeLu06}.  The
ferrofluid is filled into a cylindrical plexiglass sample holder
with inside radius $R=3.2$ mm. This radius is so small that for our
rotation frequencies the ferrofluid rotates as a rigid body with a
flow field ${\bf u}({\bf r})={\bf \Omega} \times {\bf r} = \Omega r
{\bf e}_{\varphi}$. Here $\Omega$ is the externally enforced
constant rotation rate of the sample and ${\bf e}_{\varphi}$ is the
unit vector in azimuthal direction. A homogeneous and temporally
constant magnetic field ${\bf{H}}_0=H_0{\bf{e}}_x$ is applied
perpendicular to the cylinder axis ${\bf e}_z$. For such a
combination of enforced rotation and applied field theoretical
models allow for a spatially and temporally constant nonequilibrium
magnetization ${\bf M}$ that is pushed out of the directions of
${\bf H}_0$ and ${\bf H}$ by the flow.

According to the Maxwell equations the fields ${\bf H}$ and ${\bf
M}$ within the ferrofluid are related to each other via
\begin{eqnarray}
{\bf H} = {\bf H}_0 - \frac{1}{2} {\bf M} \label{EQ:maxwell}
\end{eqnarray}
for our long cylindrical sample and in particular $H_y=-M_y/2$ as
indicated schematically in Fig.~\ref{FIG:sys}. In addition they
demand that the magnetic field outside the ferrofluid cylinder
\begin{eqnarray}
{\bf H}^{out} = {\bf H}_0 + \frac{1}{2} \frac{R^{2}}{r^{2}} \left(
2\frac{{\bf r}}{r} \frac{{\bf M}\cdot {\bf r}}{r} - {\bf M} \right)
\label{EQ:hout}
\end{eqnarray}
is a superposition of the applied field ${\bf H}_0$ and the dipolar
contribution from ${\bf M}$. This result yields a relation between
the perpendicular component of the magnetization $M_y$ resp.~of the
internal field $H_y=-M_y/2$  and the field $H_y^{sensor}$ measured
by the Hall--sensor outside the sample as indicated in
Fig.~\ref{FIG:sys}. Considering the finite size of the Hall--sensor,
$H_y^{sensor}$ is given by
\begin{equation}\label{EQ:hysensor}
H_y^{sensor} = \frac{1}{2a} \int_{-a}^a H_y^{out} dx = -
\frac{R^2}{a^2+b^2} H_y \, .
\end{equation}
In our experimental setup $b=4.75$ mm, $R=3.2$ mm, and $a=2$ mm;
here $a$ denotes the horizontal extension of the Hall sensor. So,
$H_y^{\text{sensor}}=-0.386\,H_y$ where $H_y=-M_y/2$ is the
$y$-component of the internal magnetic field in the ferrofluid.

\section{Magnetization dynamics of a poly-disperse model}

Comparisons with experimental results showed \cite{EmMaWaKiLeLu06}
that theoretical predictions \cite{LeLu06b} based on models
\cite{Sh72,FeKr99,Fe00,Sh01a,MuLi01} with a single relaxation time
overestimate the magnitude of $H_y^{sensor}$. One reason is that
particles with different sizes and different rotational dynamics of
their magnetic moments contribute differently to the
non-equilibrium, flow-induced component $M_y$ of the magnetization
and that in particular only the magnetic moments of the larger
particles with effective Brownian relaxation dynamics may be rotated
by the flow out of the direction of the magnetic field.

Therefore, we consider here as a next step poly-disperse models with
single-particle Brownian and N\'eel relaxation dynamics for the
different particle sizes. Such models have been used \cite{LeLu06a}
to determine within a linear response analysis the
effect of polydispersity on the dynamics of a torsional ferrofluid
pendulum that was periodically forced close to resonance to undergo
small amplitude oscillations in a rigid body flow
\cite{EmMuWaKnLu00,EmMuLuKn00}.

We ignore the influence of any dipolar magnetic interaction and of
any flow induced interaction on the (rotation) dynamics of the
particles. Thereby collective, collision dominated long-range and
long-time hydrodynamic relaxation dynamics of the ensemble of
magnetic moments are discarded since only the individual relaxation
of each magnetic moment is considered
--- albeit in the collectively generated internal magnetic field
${\bf H}$.

For numerical reasons we use a discrete partition of the particle
size distribution. Then, without interaction, the magnetization of
the resulting mixture of mono-disperse ideal paramagnetic gases is
given by ${\bf M}=\sum {\bf M}_j$, where ${\bf M}_j$ denotes the
magnetization of the particles with diameter $d_j$. We assume that
each magnetic moment and with it each ${\bf M}_j$  obeys a simple
Debye relaxation dynamics that drives them in the absence of flow
towards their respective equilibrium value ${\bf M}_j^{eq} ({\bf
H})$. Then, in the stationary situation resulting from the rigid
body rotation with constant ${\bf \Omega}$ the Debye relaxation
equation for each sub magnetization is given by
\begin{equation} \label{EQ:maggl}
{\bf \Omega} \times {\bf M}_j = \frac{1}{\tau_j} [{\bf M}_j - {\bf
M}_j^{eq} ({\bf H})] \, .
\end{equation}
In the absence of interactions the equilibrium magnetization of each
species is determined by a Langevin function
\begin{equation} \label{EQ:meq}
{\bf M}_j^{eq} ({\bf H}) = \chi_j (H) {\bf H} = w_j {\cal L}\left(
\frac{\mu_0 \pi M_{sat}^{bulk}}{6k_BT} d_j^3 H  \right) \,
\frac{{\bf H}}{H} \, .
\end{equation}
Here $M_{sat}^{bulk}$ is the bulk-saturation magnetization of the
magnetic material. For the magnetic weight $w_j (d_j)$ of species
$j$ we take the experimentally determined values in the
representation (\ref{meqvonh}).

We should like to draw the attention of the reader to the fact that
the magnetization equations (\ref{EQ:maggl}) for the different
particle sizes are coupled via the internal field ${\bf H}$:
according to Maxwell's equations ${\bf H}={\bf H}_0 - \frac{1}{2}
{\bf M}$ is given in terms of the total ${\bf M}=\sum {\bf M}_j$.

\begin{figure}[htp]
\centerline{\includegraphics[width=11cm,angle=0]{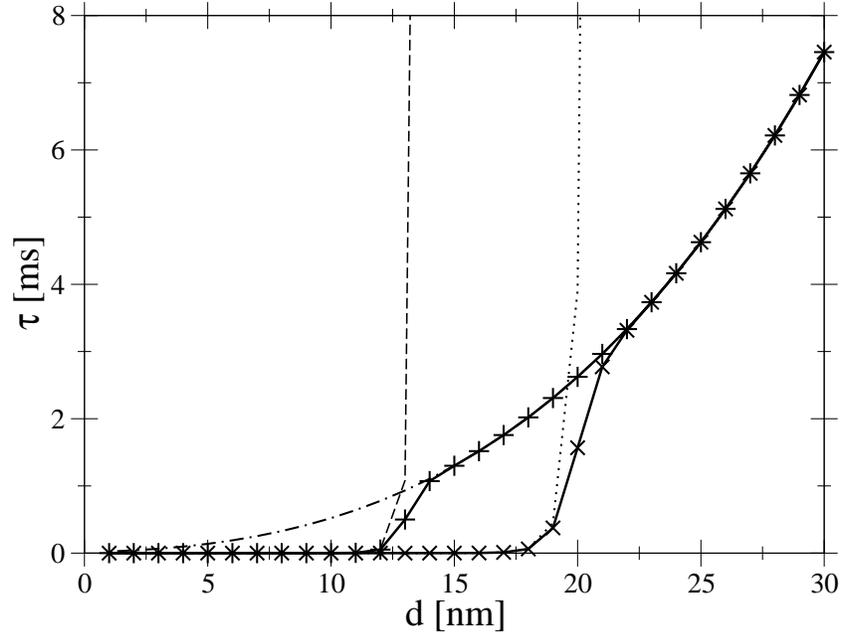}}
\caption{Relaxation times as a function of particle diameter $d$:
Brownian (dot-dashed line), N\'eel (dotted line for $K=15kJ/m^3$,
dashed line for $K=50kJ/m^3$), and the combination (\ref{EQ:tau})
(solid line with crosses for $K=15kJ/m^3$, solid line with plusses
for $K=50kJ/m^3$).} \label{FIG:tau}
\end{figure}

In the relaxation rate $1/\tau_j$ we take into account Brownian
and N\'eel relaxation processes by adding their rates with equal
weight
\begin{equation} \label{EQ:tau}
\frac{1}{\tau_j} = \frac{1}{\tau_B^j} + \frac{1}{\tau_N^j} \,.
\end{equation}
The relaxation times depend on the particle size according to
$\tau_B^j=\frac{\pi \eta}{2k_BT} (d_j+2s)^3$ and $\tau_N^j=f_0^{-1}
\exp\left(\frac{\pi Kd_j^3}{6k_BT}\right)$. Her  $\eta$ is the
viscosity, $s$ the thickness of the nonmagnetic particle layer, and
$K$ the anisotropy constant. The combined relaxation rate
(\ref{EQ:tau}) is dominated by the faster of the two processes. Thus,
large particles relax in a Brownian manner with relaxation times of
about some $10^{-3}s$, while small particles have the much smaller
N\'eel relaxation times. The boundary between N\'eel and Brownian
dominated relaxation as a function of particle size $d$ depends
sensitively on the anisotropy constant $K$. This is documented in
Fig.~\ref{FIG:tau} for the two values $K=15kJ/m^3$ and $K=50kJ/m^3$.
For these specific examples the boundaries between N\'eel and
Brownian dominated relaxation lie at about $d\simeq 20nm$ and
$d\simeq 13nm$, respectively.

\section{Comparison with experiments}
\begin{figure}[htp]
\centerline{\includegraphics[width=12cm,angle=0]{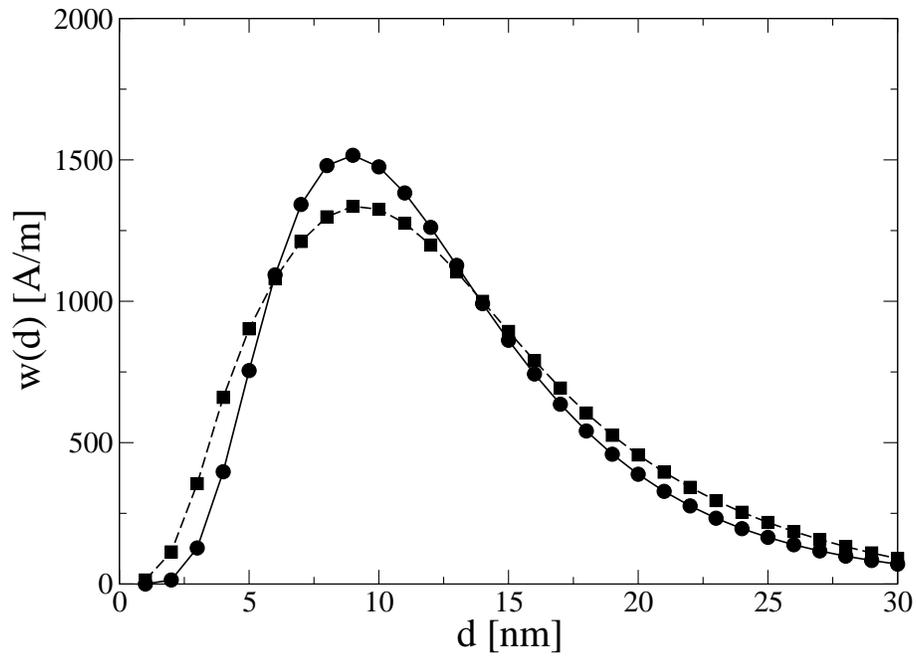}}
\caption{Magnetic weights of the 30 here considered particle sizes
($d_1=1nm$ to $d_{30}=30nm$) obtained from measurements of
$M_{eq}(H)$ in Fig.\ref{FIG:meq} by using a lognormal distribution
(solid line, circles) and by using a regularization method (dashed
line, squares).} \label{FIG:vert}
\end{figure}

For the numerical calculations we take typical values for the
ferrofluid APG 933 of FerroTec that is used among others in the
experiments described in \cite{EmMaWaKiLeLu06}: $M_{sat}^{bulk}=456
kA/m$, $\eta = 0.5 Pa\, s$, $s=2nm$, and $f_0=10^9 Hz$. Typical
values of $K$ lie between $10 kJ/m^3$ and $50 kJ/m^3$
\cite{FaPrCh99,Fa94}. We furthermore use as input the experimental
equilibrium magnetization $M_{eq}(H)$ of APG 933 shown in
Fig.~\ref{FIG:meq} and the magnetic weights of Fig.~\ref{FIG:vert}
obtained with fits to a log-normal distribution or with a
regularization method \cite{EmMaWaKiLeLu06}.

From the previous work \cite{LeLu06b,EmMaWaKiLeLu06} we know that
single-relaxation time (mono-disperse) models predict the maximum of
$M_y$ resp. of $H_y^{sensor}$ to be located roughly at $\Omega \tau
\sim 1$. Furthermore, in the experiments \cite{EmMaWaKiLeLu06} done
with poly-disperse ferrofluids for frequencies up to $\Omega \simeq
3000 rad/s$ mainly the large  particles contribute to $M_y$ resp. to
$H_y^{sensor}$ since their magnetic moments are effectively frozen
in the particle's crystal lattice. Only these magnetic moments can
be pushed out of the direction of the magnetic field by the combined
action of thermally induced rotary Brownian motion and deterministic
macroscopic flow in the rotating cylinder. Smaller particles can
keep their moment aligned with the magnetic field by the N\'eel
process when these particles undergo rotational motion. Finally, the
particle diameter that separates N\'eel behavior from Brownian
behavior in the size distribution and that thereby determines how
many particles contribute to $M_y$ resp. to the experimental
signal $H_y^{sensor}$ depends sensitively on the anisotropy constant
$K$: The smaller $K$, the smaller is the number of Brownian
particles according to Fig.~\ref{FIG:tau}, and the smaller is $M_y$
resp. $H_y^{sensor}$.

\begin{figure}[htp]
\centerline{\includegraphics[width=12cm,angle=0]{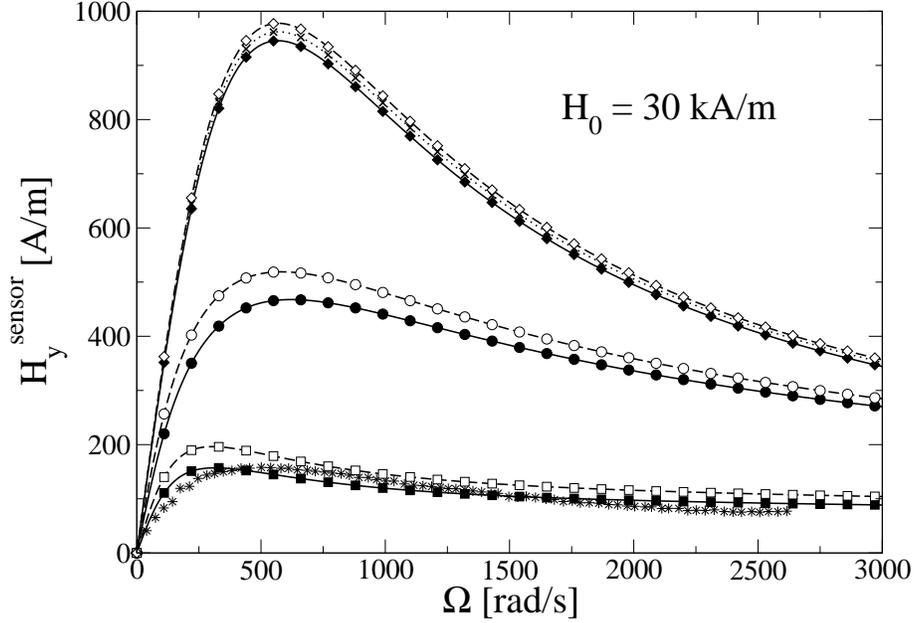}}
\caption{$H_y^{sensor}$ as function of $\Omega$ for $H_0=30 kA/m$.
The three uppermost curves refer to single-time relaxation
approximations, each with $\tau = 2ms$: monodisperse Debye model
(dotted line with crosses), polydisperse models with common $\tau_j
= \tau$ and magnetic weights $w_j$ obtained with a lognormal
distribution (full line with full diamonds) and with a
regularization method (dashed line with open diamonds). Curves with
circles and squares refer to truly polydisperse models [
eqs.~(\ref{EQ:maggl} - \ref{EQ:tau})] with $K=50 kJ/m^3$ (circles)
and $K=15 kJ/m^3$ (squares). Full lines with full symbols were
obtained with a lognormal distribution. Dashed lines with open
symbols refer to a distribution resulting from the regularization
method. In all models the equilibrium magnetization $M_{eq}(H)$ was
taken to be the experimental one.} \label{FIG:monopoly}
\end{figure}

The above sketched physical picture is corroborated by
Fig.~\ref{FIG:monopoly}. There we compare the experimentally
obtained $H_y^{sensor}(\Omega)$ (stars) as a function of $\Omega$
for a representative externally applied field $H_0=30 kA/m$ with
various model variants that take into account the polydispersity to
different degrees. This is done for two different anisotropy
constants, namely, $K=15 kJ/m^3$ and $K=50 kJ/m^3$ as representative
examples. However, the three uppermost curves refer to single time
relaxation approximations, each with $\tau = 2ms$
\cite{EmMaWaKiLeLu06}: the dotted line with crosses is the result of
a strictly monodisperse Debye model while the lines with diamonds
refer to polydisperse models, however, with common $\tau_j = \tau
=2ms$ taken in Eq.~(\ref{EQ:maggl}) but different magnetic weights
$w_j$ obtained either from a lognormal distribution (full line with
full diamonds) or from the regularization method (dashed line with
open diamonds). The equilibrium magnetization $M_{eq}(H)$ was taken
to be the experimental one, the distributions were obtained from
this experimental $M_{eq}(H)$ by the lognormal ansatz resp. the
regularization method. This, first of all, shows that models with
only one relaxation time show roughly the same behavior of
$M_y(\Omega)$ irrespective of whether the particle size and magnetic
moment distributions are polydisperse or not.

The set of curves with circles and squares in
Fig.~\ref{FIG:monopoly} refer to truly polydisperse models,
eqs.~(\ref{EQ:maggl} - \ref{EQ:tau}), but different anisotropy
constants of the magnetic material [$K=50 kJ/m^3$ (circles), $K=15
kJ/m^3$ (squares)]. Again, full lines with full symbols were
obtained with a lognormal distribution while dashed lines with open
symbols refer to a distribution resulting from the regularization
method. Here, one sees that these two distributions with their
magnetic weights displayed in Fig.~\ref{FIG:vert} yield very similar
results which might not be surprising in view of the fact that both
seem to reproduce $M_{eq}(H)$ adequately.

The largest and most important difference between the curves with
diamonds (i.e., the single-time models) and the curves with circles
and squares (i.e., the genuine polydisperse models) come from the
difference in the anisotropy constants of the magnetic material that
governs how many particles contribute efficiently as Brownian ones
to the transverse magnetization $M_y$ resp. to $H_y^{sensor}$: for
smaller $K$ the magnetic moments of fewer particles being Brownian
ones may be rotated out of the direction of the magnetic field by
the flow in the cylinder.

\begin{figure}[htp]
\centerline{\includegraphics[width=12cm,angle=0]{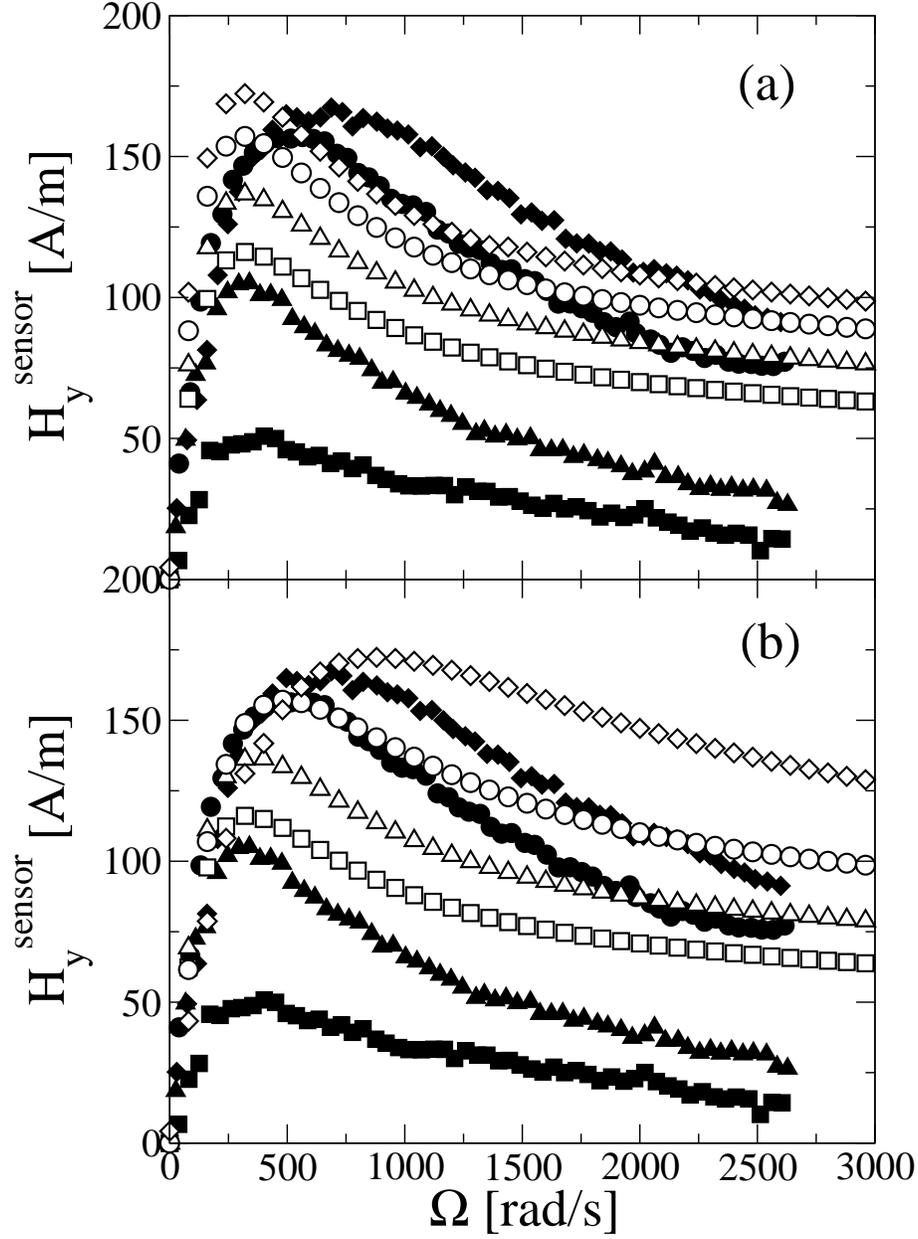}}
\caption{$H_y^{sensor}$ as function of $\Omega$ for $H_0=8.6 kA/m$
(squares), $15 kA/m$ (triangles), $30 kA/m$ (circles), and $60 kA/m$
(diamonds). Full symbols denote experimental data. Open symbols
refer to the polydisperse model [eqs.~(\ref{EQ:maggl} -
\ref{EQ:tau})] with lognormal distribution and $K=15 kJ/m^3$: in (a)
the relaxation times $\tau_j$ are independent of $H$, in (b) they
are replaced by $\tau_j^{\gamma}(H)$ (\ref{EQ:tauhgamma}) with
$\gamma = 10^{-4}m/A$. } \label{FIG:hy}
\end{figure}

\begin{figure}[htp]
\centerline{\includegraphics[width=12cm,angle=0]{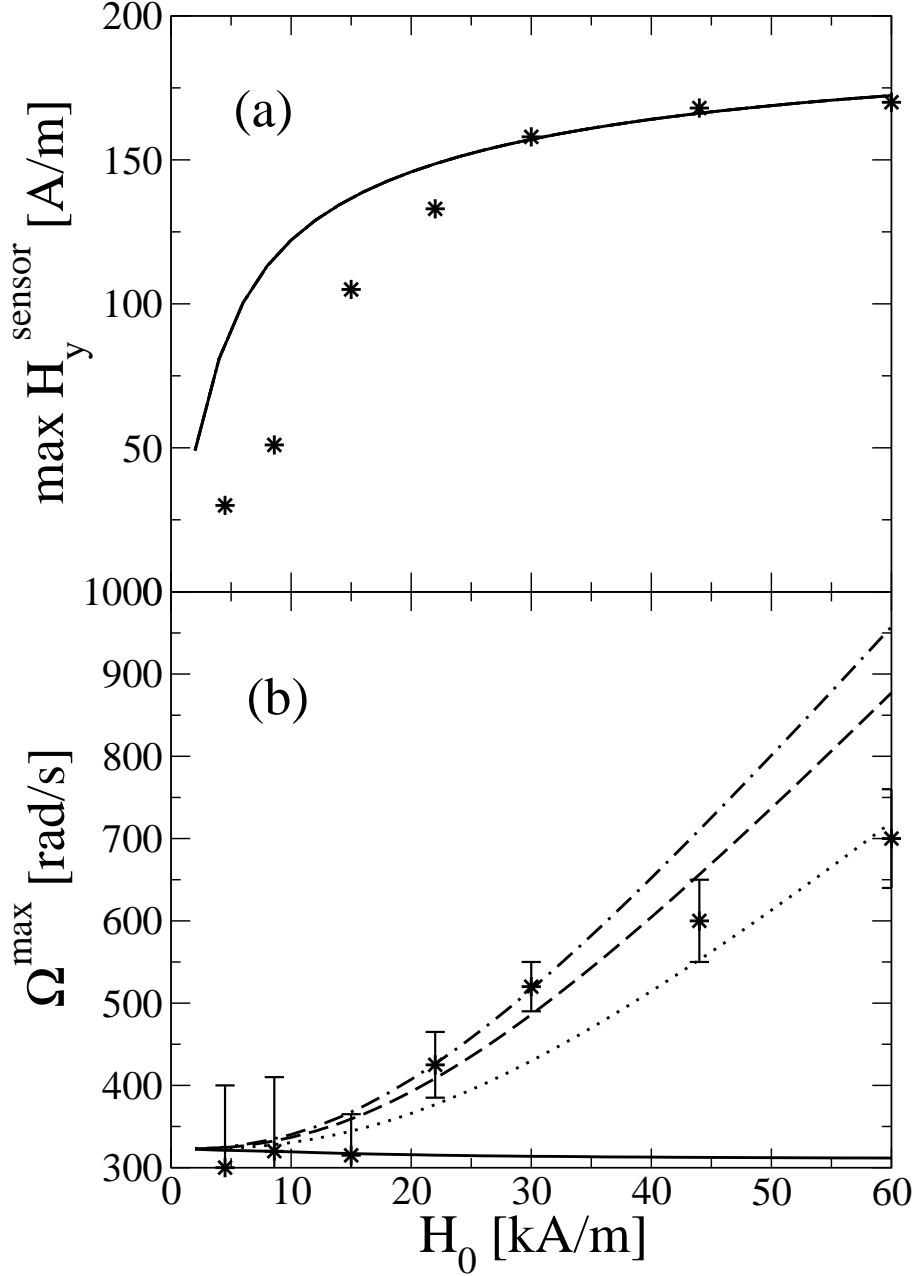}}
\caption{Maximum magnitude $max H_y^{sensor}=
H_y^{sensor}(\Omega^{max})$ (a) and location of the maximum
$\Omega^{max}$ (b) as functions of the external field $H_0$. Stars
show experimental data. Lines refer to the results of polydisperse
models with a lognormal distribution and $K=15 kJ/m^3$:
$H$-independent relaxation times (solid); $H$--dependencies with
$\gamma = 0.8\cdot10^{-4}m/A$ (dotted), $\gamma = 10^{-4}m/A$
(dashed), and $\gamma = 1.1\cdot 10^{-4}m/A$ (dot-dashed). In (a)
the differences between the lines are too small to be seen.}
\label{FIG:max}
\end{figure}

The curves for $K=15 kJ/m^3$ yield roughly the same maximal size
$H_y^{sensor}$ as the experiments --- they could be fine-tuned even
further. But then the location, $\Omega^{max}(H_0)$, of the maxima
for different $H_0$ is still off from the experimental ones as shown
in Fig.~\ref{FIG:hy}(a) and \ref{FIG:max}(b). However, the agreement
between the predictions of the polydisperse models of
eqs.~(\ref{EQ:maggl} - \ref{EQ:tau}) and the experiments concerning
the location $\Omega^{max}(H_0)$
can be improved by allowing the relaxation rates $\tau_j$ of the
differently sized particles to depend also on the internal field
$H$. To demonstate this, we use for simplicity the form
\cite{EmMaWaKiLeLu06}
\begin{equation}\label{EQ:tauhgamma}
\tau_j^{\gamma} (H) = \tau_j \, \frac{2{\cal L}(\gamma H)}{\gamma H
- {\cal L}(\gamma H)}
\end{equation}
with one additional fit parameter $\gamma$. Values of about $\gamma
\sim 10^{-4} m/A$ yield maximum locations $\Omega^{max}(H_0)$ that
agree well with the experiments as can be seen in
Fig.~\ref{FIG:hy}(b) and Fig.~\ref{FIG:max}(b). This generalization
of the model (\ref{EQ:maggl} - \ref{EQ:tau}) leaves the peak value
$H_y^{sensor}(\Omega^{max})$ almost unchanged, cf,
Fig.~\ref{FIG:max}(a).

However, also this augmented polydispersive model reproduces with
fixed values of $K$ and $\gamma$ the experimental data only in a
small range of $\Omega$ and $H_0$, cf, Fig.~\ref{FIG:hy} and
Fig.~\ref{FIG:max}.

\section{Conclusion}
We have compared the predictions of polydisperse models of the
magnetization dynamics of ferrofluids with recent
experiments measuring the transverse magnetization component $M_y$
of a rotating ferrofluid cylinder. The models use mixtures of
mono-disperse ideal paramagnetic gases of differently sized
particles. The  magnetization dynamics of the models take into
account the rigid body rotation of the fluid combined with a simple
Debye relaxation of the magnetic moments of each particle with size
dependent Brownian and N\'eel magnetic relaxation times. Thus, in
the absence of flow, each magnetic moment and with it each
sub-magnetization would be driven independently of the others
towards its respective mean equilibrium value that, however, depends
on the internal magnetic field ${\bf H}$ being collectively
generated by all magnetic moments.

The comparison suggests that mainly the large  particles contribute
to $M_y$ since their magnetic moments are effectively frozen in the
particle's crystal lattice. Only these magnetic moments can be
pushed effectively out of the direction of the magnetic field by the
combined action of thermally induced rotary Brownian motion and
deterministic macroscopic flow in the rotating cylinder. Smaller
particles can keep their moment aligned with the magnetic field by
the N\'eel process when these particles undergo rotational motion.

Finally, the particle diameter that separates N\'eel behavior from
Brownian behavior in the size distribution and that thereby
determines how many particles contribute to $M_y$ resp. to the
experimental signal $H_y^{sensor}$ depends quite sensitively on the
anisotropy constant $K$ of the magnetic material. $K$ determines how
many magnetic moments are "frozen" or "blocked" in particles and
thus can be rotated by the rigid body flow: The smaller $K$, the
smaller is the number of Brownian particles with frozen moments, and
the smaller is the resulting $M_y$. Or, vice versa, a large
transverse magnetization can be expected in ferrofluids with large
anisotropy constants.

An analysis of the rotation rates $\Omega^{max}(H_0)$ for which
$M_y$ is largest indicates that the agreement between experiments
and model predictions can be improved by allowing the relaxation
rates of the differently sized particles to depend also on the
magnetic field $H$.

\begin{acknowledgments}
This work was supported by the DFG (SFB 277) and by INTAS(03-51-6064).
\end{acknowledgments}

\clearpage



\clearpage

\clearpage

\clearpage

\clearpage

\clearpage

\clearpage

\clearpage

\end{document}